\documentclass[12pt]{article}
\usepackage{geometry,enumerate,amsmath,amssymb}
\usepackage{fullpage}
\usepackage{graphicx}
\newcommand{\be}{\begin{equation}}
\newcommand{\ee}{\end{equation}}

\renewcommand{\epsilon}{\varepsilon}

\newcommand{\news}{\setcounter{equation}{0}\quad}
\def\ben{\begin{equation}}
\def\een{\end{equation}}
\def\bea{\begin{eqnarray}}
\def\eea{\end{eqnarray}}
\begin{document}
\title{
\begin{flushright}\ \vskip -2cm {\small {\em DCPT-12/17}}\end{flushright}
\vskip 2cm Hyperbolic vortices with large magnetic flux}
\author{
Paul Sutcliffe\\[10pt]
{\em \normalsize Department of Mathematical Sciences, Durham University, Durham DH1 3LE, U.K.}\\[10pt] {\normalsize Email:  \quad p.m.sutcliffe@durham.ac.uk} 
}
\date{April 2012}
\maketitle
\begin{abstract}
There has been some recent interest in the study of 
non-abelian BPS monopoles in the
limit of large magnetic charge.
Most investigations have used a magnetic bag
approximation, in which spherical symmetry is assumed within an 
abelian description. In particular, this approach has been used to
suggest the existence of two types of magnetic bag,
with differing distributions of the zeros of the Higgs field,
together with multi-layer structures, containing several
magnetic bags. This paper is concerned with the analogous situation of 
abelian BPS vortices in the hyperbolic plane, in the limit
of large magnetic flux. This system has the advantage that explicit 
exact solutions can be obtained and compared with a magnetic bag 
 approximation. Exact BPS vortex solutions are presented that 
are analogous to the two types of magnetic bags predicted for BPS monopoles
and it is shown that these structures can be combined to produce 
exact multi-layer solutions.
\end{abstract}
\newpage
\section{Introduction}\news
The concept of a magnetic bag was first introduced by Bolognesi \cite{Bo1} in
the context of vortices in the abelian Higgs model in the Euclidean plane,
where it is an approximation to a vortex solution with a large magnetic
flux. In the simplest case the magnetic bag has circular symmetry and
the approximation assumes that inside the bag the magnetic field is
constant and the Higgs field vanishes, whereas outside the bag the magnetic
field vanishes and the Higgs field takes its vacuum expectation value.
Neither the magnetic field nor the Higgs field are continuous at the surface
of the bag, but the surface contribution to the energy can be neglected 
because it is subleading in
the limit of a large vortex number  $N\gg1.$ A comparison with 
numerical solutions of
circularly symmetric vortices 
provides support for the validity of this
approximation \cite{BG}. Magnetic bags are particularly interesting in
the BPS limit of critically coupled vortices, 
where it is expected that the shape of the bag can be varied because of 
the existence of the $2N$-dimensional moduli space of BPS 
vortex solutions.

Bolognesi \cite{Bo3} extended the magnetic bag idea to monopoles in
non-abelian Yang-Mills-Higgs theories in 
three-dimensional Euclidean space, 
where they describe configurations with large magnetic charge.
Again the most interesting situation is the BPS limit, in this case associated 
with a massless Higgs field. In this context the approximation assumes that 
inside the bag both the Higgs field and the magnetic field vanish, whereas 
outside the bag the fields are taken to be abelian, consisting of a magnetic 
field and a scalar field that represents the magnitude of the non-abelian 
Higgs field. These abelian fields satisfy the abelianized version of the 
Bogomolny equation. 
The simplest situation is to assume that the bag is spherically symmetric, 
though again it is expected that the shape of the bag can be varied because
of the $4N$-dimensional moduli space associated with an $N$-monopole solution.

The magnetic bag description has been applied to several different
aspects of monopoles and used to predict a number of interesting phenomena. 
Lee and Weinberg \cite{LW} have proposed that there are a range
of magnetic bags, all of which have approximate spherical symmetry, but are
distinguished by the distribution of the zeros of the Higgs field associated
with the exact monopole solution. In particular, they propose that there is
an extreme case, in which all the Higgs zeros are coincident at the 
centre of the bag, and a second extreme case in which the Higgs zeros are 
distinct and (almost all) are located near the surface of the bag.
Manton \cite{Ma} has investigated the properties of multi-layer structures, 
created by patching together nested sequences of magnetic bags, and
Harland \cite{Ha} has shown how magnetic bags may be described using a
large $N$ limit of the Nahm transform. The magnetic bag approximation
has also proved useful in the study of monopoles in Anti-de Sitter
spacetime \cite{BT}, where it compares well with numerical solutions \cite{Su} 
even for rather small values of $N.$
 
The purpose of the present paper is to investigate some of the
above issues in a system where exact explicit solutions are
available for comparison with the magnetic bag approximation.
The chosen system is the abelian Higgs model in the hyperbolic plane
and its associated BPS vortices. 
In particular, 
solutions with large magnetic flux are studied that include 
analogues of the two extreme types of magnetic bags predicted for BPS monopoles
and it is shown that these structures can be combined to produce 
exact multi-layer solutions.

\section{Hyperbolic vortices}\news
This section contains a brief review of BPS vortices in the hyperbolic 
plane. A more detailed discussion of this material can be found in chapter 7 
of \cite{MSbook}.

The system of interest is the abelian Higgs model, with complex scalar
Higgs field $\phi$ and real gauge potential $a_i.$ 
The theory is defined on the hyperbolic plane ${\cal H}^2$ of curvature
$-\frac{1}{2},$ with metric
\be
ds^2=\Omega(dx^2+dy^2),
 \quad \mbox{where}\quad
\Omega=\frac{8}{(1-x^2-y^2)^2}
\ee
and $(x,y)$ are coordinates in the Poincar\'{e} disc model of
 ${\cal H}^2,$ with $x^2+y^2< 1.$
The energy of the model is given by the standard Ginzburg-Landau energy
at critical coupling
\be
E=\int_ {{\cal H}^2}\bigg(
\frac{1}{2}\Omega^{-1}B^2
+\frac{1}{2}D_i\phi\overline{D_i\phi}
+\frac{\Omega}{8}(1-|\phi|^2)^2
\bigg)\ d^2x,
\label{energy}
\ee
where $B=\partial_1 a_2-\partial_2 a_1$ is the magnetic field and 
$D_i\phi=\partial_i\phi-ia_i\phi$ the covariant derivative of the
Higgs field.
The vacuum expectation value
of the Higgs field has been set to unity for convenience.

The magnetic flux through the hyperbolic plane is quantized,
\be
\int_ {{\cal H}^2} B\,d^2x
=2\pi N,
\ee
where the integer $N$ is the vortex number and is equal to the
winding number of the phase of the Higgs field on the circle at infinity.
$N$ is the first Chern number of the gauge field 
and it is also equal to the number of
zeros of the Higgs field, counted with multiplicity.
Without loss of generality, the vortex number $N$ will be 
taken to be positive in this paper.

A standard Bogomolny argument yields the energy bound $E\ge \pi N,$
with equality attained by the BPS vortex solutions that solve
the first order Bogomolny equations
\be
D_1\phi+iD_2\phi=0, \quad \mbox{and}\quad B=\frac{\Omega}{2}(1-|\phi|^2).
\label{bog}
\ee
As first observed by Witten \cite{Wi2}, these Bogomolny equations
are integrable for a particular value of the curvature of the hyperbolic
plane ($-\frac{1}{2}$ in the conventions of this paper, hence its
adoption from the outset). 

Explicitly, the general $N$-vortex solution can be
obtained in closed form by introducing the complex coordinate 
$z=x+iy,$ 
in the unit disc, and setting
\be
\phi=\bigg(\frac{1-|z|^2}{1-|f|^2}\bigg)
\frac{df}{dz}
\quad \mbox{and} \quad 
a_z=-i\frac{\partial}{\partial \bar z}\log
\bigg(\frac{1-|z|^2}{1-|f|^2}\bigg),
\label{gensoln}
\ee
where $f(z)$ is a rational map of the form
\be
f=z\prod_{i=1}^N\bigg(
\frac{z-\beta_i}{1-\bar \beta_iz}\bigg),
\ee
with complex constants $\beta_i$, all inside the unit disc $|\beta_i|<1.$
The gauge freedom can be used to write any $N$-vortex solution 
in this form.

The $N$ complex parameters $\beta_i$ provide coordinates for the $N$-vortex
moduli space ${\cal M}_N,$ which has real dimension $2N.$ These
parameters determine the positions of the $N$ vortices, which are
given by the points in the unit disc where 
$\frac{df}{dz}$ 
vanishes,
since it follows from (\ref{gensoln}) that  
$\phi$ is zero at these points. In general, the relation
between the solution parameters $\beta_i$ and the vortex positions is
only known implicitly, since the relation requires knowledge of the
zeros of the derivative of $f(z).$

The simplest example of an $N$-vortex solution represents 
$N$ coincident vortices at the origin and is given by the
map $f=z^{N+1},$ corresponding to the choice $\beta_i=0$ $\forall i.$
This solution has circular symmetry, since under an arbitrary 
spatial rotation through an angle $\chi$, given by $z\mapsto ze^{i\chi},$ 
the map transforms as $f\mapsto fe^{i(N+1)\chi},$ and a change in the phase
of $f$ is simply a gauge transformation. In particular, the modulus
of the Higgs field depends only on the distance from the origin and
is given by
\be
|\phi|=\frac{(N+1)|z|^N}{|z|^{2N}+|z|^{2N-2}+\cdots+1}.
\label{axial}
\ee

\section{Vortices as magnetic bags}\news
A magnetic bag description of hyperbolic vortices 
in the large $N$ limit
can be obtained by making the approximation 
that the Higgs field vanishes identically throughout some 
region 
${\cal R}\subset {\cal H}^2,$ with area ${\cal A}.$ 
In this region the Bogomolny equations (\ref{bog}) then imply that
$B=\Omega/2,$ that is, there is a constant magnetic flux per unit area.
Outside the region ${\cal R}$ the Higgs field is taken to have its 
vacuum value $|\phi|=1,$ and the Bogomolny equations then determine
that $B=0$ in this region of space. 
If the gauge is chosen so that $\phi$ is real outside ${\cal R}$
then $\phi=1$ and the energy density vanishes identically in this 
region. Generally the gauge in which $\phi$ is real is not a good one
since an $N$-vortex configuration has the phase of $\phi$  winding 
$N$ times on the circle at infinity, so a gauge in which $\phi$ is real 
must be singular. However, it is assumed that the singularities occur
inside the region ${\cal R},$ or on its boundary $\partial {\cal R},$
where this approximation for the Higgs field does not apply.

In this magnetic bag approximation the Higgs field and the magnetic field 
are not continuous on the boundary 
$\partial {\cal R},$ but neglecting the boundary contribution to the
energy, which should be a reasonable approximation in the large $N$ limit,
the energy (\ref{energy}) becomes
\be
E=\int_{\cal R}\frac{1}{4}\Omega
\,d^2x
=\frac{1}{4}{\cal A}.
\ee
Within this approximation the vortex number is given by
\be
N=\frac{1}{2\pi}\int_{\cal R} B \,d^2x=\frac{1}{4\pi}\int_{\cal R} \Omega \,d^2x
=\frac{1}{4\pi}{\cal A}
\ee
and combining these two expressions yields the BPS energy formula 
$E=\pi N.$

The validity of this magnetic bag approximation can be tested in
the simple case of a circular bag by comparing with the 
exact $N$-vortex solution (\ref{axial}).
The Poincar\'{e} disc coordinate $z$ is not very convenient for
analysing vortices in the large $N$ limit, because its modulus
 has a finite range. Therefore a coordinate transformation is
first made by writing $z=e^{i\theta}\tanh(r/2^{3/2})$ so that
the metric on the hyperbolic plane becomes
 \be
ds^2=dr^2+2\sinh^2(r/\sqrt{2})\,d\theta^2,
\label{metric}
\ee
where the radius $r\in[0,\infty)$ is the geodesic distance to the origin.
In terms of this coordinate, the circular solution (\ref{axial})
becomes
\be
|\phi|=\frac{2(N+1)(e^{\sqrt{2}r}-1)^N}{\sum_{j=0}^N
\binom{2N+2}{2j+1}e^{\sqrt{2}jr}}.
\label{circular}
\ee
In particular, the 1-vortex takes the simple kink form
$|\phi|=\tanh(r/\sqrt{2}).$

The vortex number of a circular magnetic bag of radius $r=R$ is 
given by
\be
N=\frac{1}{4\pi}{\cal A}=\cosh(R/\sqrt{2})-1,
\ee
so in the large $N$ limit,
and neglecting terms that decay with $N,$
the expression for the radius is 
\be
R=\sqrt{2}\log(2N).
\label{bigr}
\ee
\begin{figure}[ht]\begin{center}
\includegraphics[width=12cm]{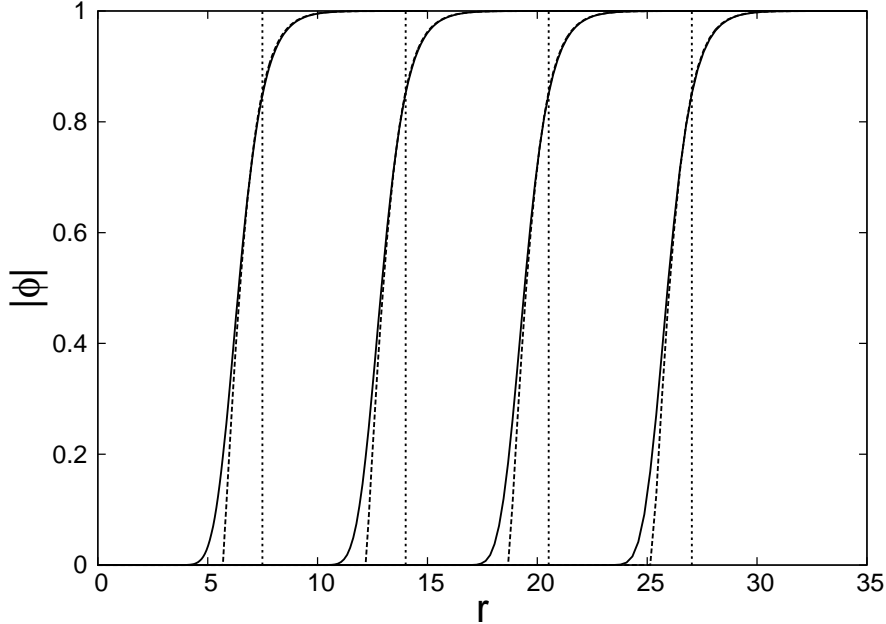}
\caption{The modulus of the Higgs field $|\phi|$ as a function of
the radius $r,$ for circularly symmetric vortices with vortex numbers
$N=10^2,10^4,10^6,10^8$ (curves move to the right with increasing $N$).
Solid curves represent the exact solution (\ref{circular}).
Vertical dashed lines denote the value of the bag radius $R$ given
by (\ref{bigr}). Dashed curves correspond to the improved magnetic
bag approximation (\ref{newbag}).
}
\label{fig-radial}\end{center}\end{figure}
In Figure~\ref{fig-radial} the solid curves represent the exact
solution (\ref{circular}) with $N=10^2,10^4,10^6,10^8$ (curves move
to the right with increasing $N$). The dashed vertical lines correspond
to the associated magnetic bag radius (\ref{bigr}), revealing a reasonable
approximation to the size of the vortex. However, the 
surface of the bag is assumed to have zero thickness, when in fact
it is determined by the Higgs mass and has a finite value that is independent
of $N.$ Therefore a better approximation is to assume the same thickness
as in the exact 1-vortex solution and take the continuous function
\be
|\phi|=\tanh((r-R_0)/\sqrt{2})\Theta(r-R_0),
\label{newbag}
\ee
where $\Theta$ is the Heaviside step function and
$R_0$ is a parameter that may be viewed as the radius of the magnetic 
bag with a finite thickness.
The relation between the radius $R_0,$ and the vortex number $N$ is
determined by substituting the expression (\ref{newbag}) into the formula
\be
N=\frac{1}{2\pi}\int B\,d^2x
=\frac{1}{4\pi}\int (1-|\phi|^2)\,d^2x.
\ee
Neglecting terms that tend to zero as $N\rightarrow\infty,$ 
this formula may be written as
\be
R_0=R-\sqrt{2}\log\bigg(\frac{16-\pi^2}{8-2\pi}\bigg),
\label{r0}
\ee
which confirms that $R_0$ differs from the naive radius $R$ only by
a term that is $O(1),$ associated with the finite thickness of the 
magnetic bag surface. The dashed curves in Figure~\ref{fig-radial} 
correspond to the improved magnetic bag approximation (\ref{newbag})
with the radius given by (\ref{r0}). It is clear that this yields
an excellent approximation to the exact $N$-vortex solution.

The kink that appears as the large $N$ boundary bag profile is the
hyperbolic vortex analogue of the monopole wall \cite{Wa}.
Its form can be obtained\footnote{I thank Nick Manton for this observation}
by making a connection to the work of Manton and Rink \cite{MR}
on a single vortex on a hyperbolic trumpet. Identify the angle $\theta$
with its shift by $2\pi/N,$ to give a hyperbolic cone containing a single
vortex. Introduce the angle on the cone $\chi_1=N\theta,$ 
which has period $2\pi,$ and
make a change of variable from $r$ to $\chi_2$ by writing 
$r=R-\sqrt{2}\log\chi_2,$ where $R$ is given by (\ref{bigr}).
In terms of the coordinates $\chi_1,\chi_2$ the metric (\ref{metric})
becomes 
\be
ds^2=\frac{2}{\chi_2^2}(d\chi_1^2+d\chi_2^2),
\label{uhp}
\ee
where the large $N$ limit has been used to replace the hyperbolic
function in (\ref{metric}) by its exponential approximation for
large argument. The metric (\ref{uhp}) is the upper half plane model
of hyperbolic space, but as $\chi_1$ is periodic this surface is a
hyperbolic trumpet. The vortex on this surface has been obtained in
\cite{MR} and yields
$|\phi|=\chi_2/\sinh\chi_2.$
Converting back to the radial variable $r$ the boundary bag profile is
\be
|\phi|=\frac{e^{(R-r)/\sqrt{2}}}{\sinh(e^{(R-r)/\sqrt{2}})}
=\frac{2Ne^{-r/\sqrt{2}}}{\sinh(2Ne^{-r/\sqrt{2}})},
\label{profile}
\ee
which has a novel double exponential form. 

\begin{figure}[ht]\begin{center}
\includegraphics[width=12cm]{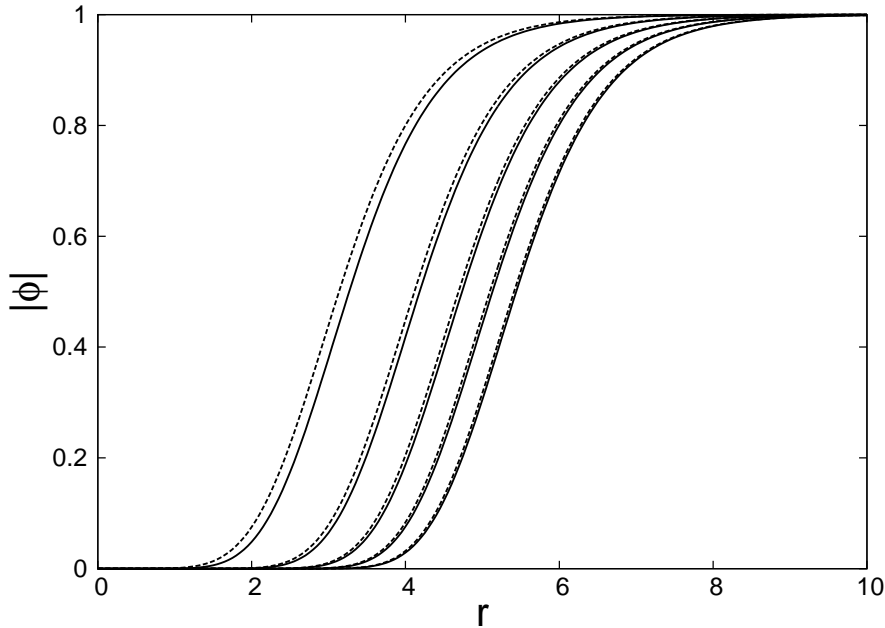}
\caption{The modulus of the Higgs field $|\phi|$ as a function of
the radius $r,$ for circularly symmetric vortices with vortex numbers
$N=10,20,30,40,50$ (curves move to the right with increasing $N$).
Solid curves represent the exact solution (\ref{circular})
and dashed curves correspond to the 
boundary bag profile (\ref{profile}).
}
\label{fig-trumpet}\end{center}\end{figure}

Although the boundary bag profile (\ref{profile}) has been derived in
the large $N$ limit, it provides an excellent approximation to the 
exact $N$-vortex solution (\ref{circular}) even for reasonably small
values of $N.$ This is demonstrated in Figure~\ref{fig-trumpet}, where
solid curves represent the exact solution and dashed curves the
boundary bag profile for $N=10,20,30,40,50.$

Previous studies of magnetic bags, for both vortices and monopoles,
have often used length units that scale with the bag radius,
so that the bag position remains fixed as $N$ increases, 
to facilitate the comparison of bags with different values of $N.$ 
Applied to the current situation an appropriate
scaled radial variable is $r/R,$ in terms of which the Higgs field
of the exact $N$-vortex solution approaches the magnetic bag
step function in the large $N$ limit, because of the logarithmic growth 
of $R$ with $N$ and the finite thickness of the bag. 

\section{Deforming the magnetic bag}\news
Magnetic bags assume that the Higgs field vanishes identically in a
large region of space, but for a finite number of either monopoles or 
vortices there are only a finite number of points in space where the
Higgs field is zero. In the case of monopoles, 
Lee and Weinberg \cite{LW} have proposed  that there are two extreme types 
of magnetic bag, 
characterized by different distributions of the Higgs zeros.
Both types of magnetic bags are approximately spherical, 
even though multi-monopoles with exact spherical symmetry do not exist
in the $SU(2)$ Yang-Mills-Higgs gauge theory. In the first type of
magnetic bag, termed a non-abelian bag in \cite{LW}, 
the Higgs zeros are coincident at the centre of the bag. 
The hyperbolic vortex analogue of this magnetic bag is 
clearly the circularly symmetric solution discussed in the previous section,
where all $N$ zeros of the Higgs field are located at the origin.
The term non-abelian is not appropriate in the vortex context, so this type
of solution will be referred to as a core magnetic bag, to denote that
the zeros of the Higgs field lie deep within the core of the vortex.
In the second type of magnetic bag, called an abelian bag in the monopole
context \cite{LW}, the zeros of the Higgs field are distinct and most are
located near the surface of the bag. The idea 
is that the zeros of the Higgs field may be 
viewed as moduli for the magnetic bag and
varying the Higgs zeros inside the bag has little impact on the surface
of the bag, but does change the bags character, and in particular the nature
of the fields inside the bag. The evidence in support of this view 
\cite{LW} consists of an analysis of the topological features of the 
Higgs zeros, together with an extrapolation based on 
explicit knowledge of the Higgs zeros for some low charge monopoles 
with Platonic symmetry. 

In this section an analysis will be presented of exact BPS vortex solutions 
that have approximate circular symmetry and are obtained as deformations
of core magnetic bags, in which the Higgs zeros move out from the centre
towards the surface of the bag. This provides an explicit realization
for hyperbolic vortices of the predicted behaviour for monopoles and
yields vortex analogues of both abelian and non-abelian monopole magnetic
bags. 

Consider the exact BPS $N$-vortex solution
generated from the function
\be
f(z)=\frac{z(z^N-\alpha)}{1-\alpha z^N}
\label{fcyclic}
\ee
using the formula (\ref{gensoln}). Here $\alpha\in[0,1)$ is a real 
parameter that controls the separation of the Higgs zeros.
If $\alpha=0$ then this solution reverts to the circularly symmetric
solution of the previous section, that is, a core magnetic bag
with $N$ coincident Higgs zeros at the origin.
If $\alpha>0$ then the solution has a cyclic $C_N$ symmetry and the
$N$ zeros of the Higgs field are given by $z=\rho e^{2\pi ik/N},$ where
$k=0,1,\ldots,N-1$ and $\rho\in(0,1)$ is the real root of the equation
\be
\rho^{2N}-\rho^N\bigg(N(\alpha^{-1}-\alpha)+\alpha^{-1}+\alpha\bigg)+1=0.
\label{roots}
\ee
In addition to determining the separation of the Higgs zeros,  
the parameter $\alpha$ also controls the nature of the Higgs field in the
core of the bag, since at the origin ($z=0$) the formula (\ref{gensoln})
gives that $|\phi|=\alpha.$ 
If $\alpha\approx 0$ then the Higgs zeros are close to coincidence and
the solution is only a small perturbation of the core magnetic bag.
To obtain a bag in which the Higgs zeros are close to the surface of the 
bag requires the opposite limit,
that is $\alpha\approx 1.$ However, this regime also includes a range
of $\alpha$ for which the solution (\ref{fcyclic}) describes $N$ well-separated
single vortices located on the vertices of a regular $N$-gon and
this is not a magnetic bag. 

To determine an appropriate range
of $\alpha$ it is useful to exchange $\alpha$ for the magnetic
bag deformation parameter $p\in[0,\infty)$ defined by
\be
\alpha=1-N^{-p}.
\label{p}
\ee
If $p=0$ then the circularly symmetric solution is recovered, 
so the core magnetic bag is undeformed.
Now consider the case $p>0,$ so that $\alpha\rightarrow 1$ as
$N\rightarrow \infty.$ 
The crucial issue is whether the Higgs zeros remain inside the 
magnetic bag as $p$ is increased from zero.
Recall that the relation between $|z|$ and the radial coordinate $r$ is
given by $|z|=\tanh(r/2^{3/2})\approx 1-2e^{-r/\sqrt{2}},$ in the large $N$ limit.
The circle associated with the surface of the bag has a 
radius $r=R=\sqrt{2}\log(2N)$ and corresponds to a value 
of $|z|=\rho_\star$ given by $\rho_\star=1-\frac{1}{N},$
which means that $\rho_\star^N\rightarrow 1/e$ as $N\rightarrow \infty.$
The important point about this result is that as $N\rightarrow \infty$
then $\rho_\star^N$ has a non-zero limit that is stricty less than one.
This is the property required by $\rho,$ the solution of equation
(\ref{roots}), to make sure that the $C_N$ symmetric solution does not
describe $N$ well-separated vortices. 

Note that if there are $N$ 
equally spaced vortices on a circle with a radius equal to that of the 
corresponding bag $r=R,$ then the distance between neighbouring vortices is 
$O(1),$ which is the same order as the thickness of the bag surface.
Although the vortices are not well-separated in this regime, 
they are dilute, in the sense that a circularly symmetric 
magnetic bag approximation is only valid as an averaged description, 
since there is a large angular variation in $|\phi|.$ 
This will be demonstrated below by considering the angular average
\be 
\langle|\phi|\rangle=\frac{1}{2\pi}\int_0^{2\pi} |\phi| \,d\theta
\label{average}
\ee
as a function of the radius $r.$

Substituting the form $(\ref{p})$ into the root equation (\ref{roots}) gives
\be
\rho^{2N}-2(1+N^{1-p})\rho^N +1=0,
\ee
so clearly the critical value is $p=1.$
For $p=0$ a core magnetic bag is obtained and as $p\rightarrow 1$ the
dilute regime emerges. The bag deformation parameter $p\in[0,1]$
describes the transition between these two extreme regimes.  

\begin{figure}[ht]\begin{center}
\includegraphics[width=17cm]{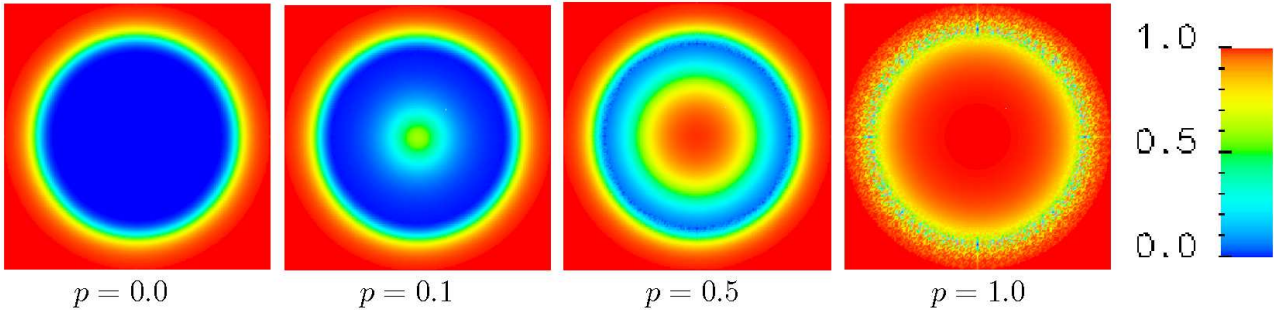}
\caption{The modulus of the Higgs field, $|\phi|,$ 
for a vortex solution with $N=10^4$ and increasing bag parameter 
$p=0.0,\,0.1,\,0.5,\,1.0.$
}
\label{fig-varyp}\end{center}\end{figure}

\begin{figure}[ht]\begin{center}
\includegraphics[width=10cm]{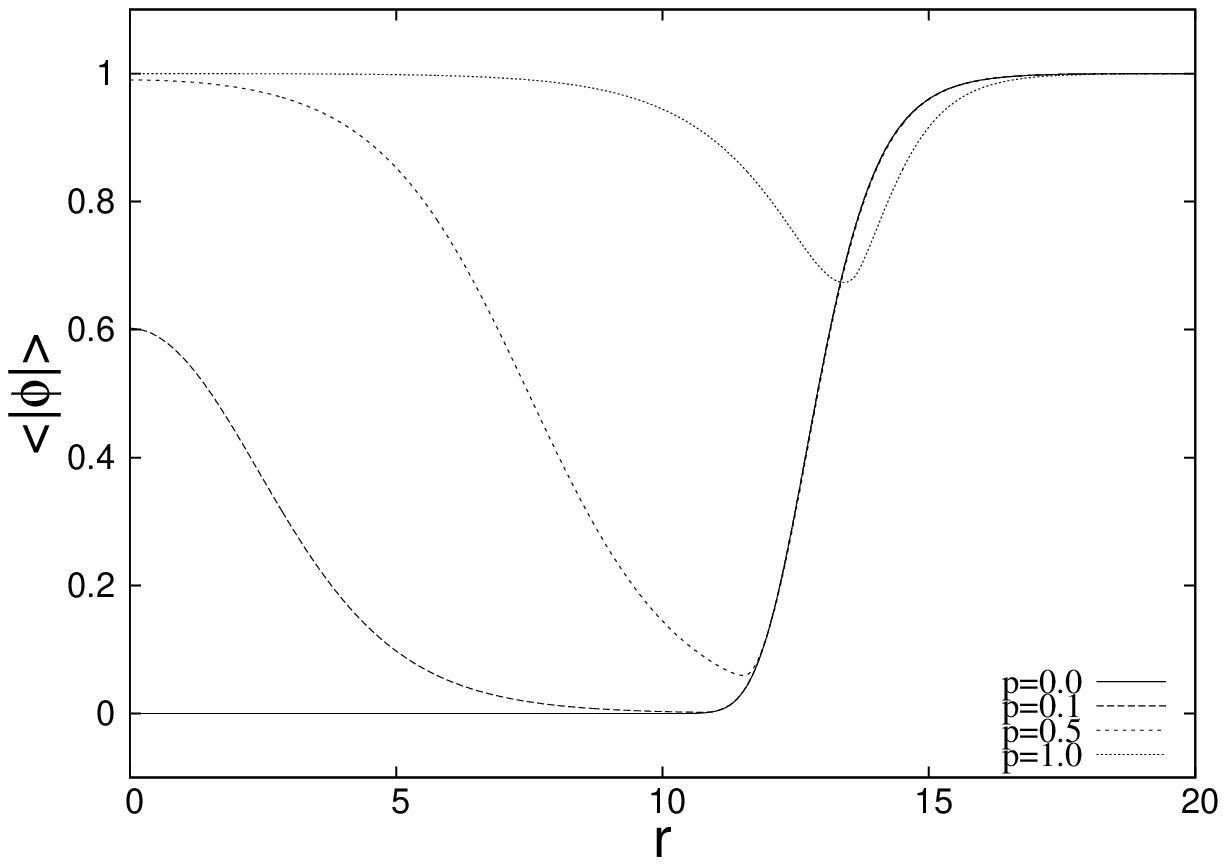}
\caption{The angular average of the modulus of the Higgs field, 
$\langle|\phi|\rangle,$ as a function of the radius $r,$ 
for a vortex solution with $N=10^4$ and increasing bag parameter 
$p=0.0,\,0.1,\,0.5,\,1.0.$
}
\label{fig-aver}\end{center}\end{figure}

To illustrate this behaviour, Figure~\ref{fig-varyp} displays 
$|\phi|$ for a solution with vortex number $N=10^4$ and increasing values
of the bag deformation parameter $p$ from $0$ to $1$.
The plotted region corresponds to $-1.2R\le X,Y \le 1.2R,$ where
$X,Y$ are the Cartesian coordinates given by $X+iY=r e^{i\theta},$ and
$R=\sqrt{2}\log(2N)$ is the bag radius. These plots confirm the expected
behaviour, with little variation of the surface of the bag, until
the dilute regime is obtained, but a significant change in the
character of the field in the interior of the bag, as it changes from
a region of unbroken symmetry to a region of broken symmetry where the
Higgs field attains its vacuum expectation value.

In Figure~\ref{fig-aver} the angular average (\ref{average}) is plotted
as a function of the radius $r$ for the four solutions displayed in
Figure~\ref{fig-varyp}. Recall that the modulus of the Higgs field at
the origin increases with $p$ since it is given by $|\phi|=1-N^{-p}.$
It is clear from these graphs that the surface of the bag is virtually
identical for the first three values $p=0.0,\,0.1,\,0.5,$ even though the
interior of the bag changes dramatically. For $p=0.1$ and $p=0.5$ the
minimum value of the angular average $\langle|\phi|\rangle$ is close to 
zero and is attained near the surface of the bag. For this range of 
$p$ there are angular directions along which the Higgs field 
exactly vanishes at a radius near to the surface of the bag, so the fact
that the angular average is also close to zero confirms the 
approximate circular symmetry of the solution that arises when the 
vortices are in a dense regime. However, in the dilute regime $p=1.0$
the angular average does not deviate far from the vacuum expectation
value, despite the fact that there are still angular directions along 
which the Higgs field exactly vanishes at a radius near to the 
surface of the bag. In this sense the solution is far from being
circularly symmetric and there is a large angular variation of the fields.
In this dilute regime it is clear that any attempted magnetic bag 
approximation can only be valid in terms of a description of angularly 
averaged fields.   

\begin{figure}[ht]\begin{center}
\includegraphics[width=8cm]{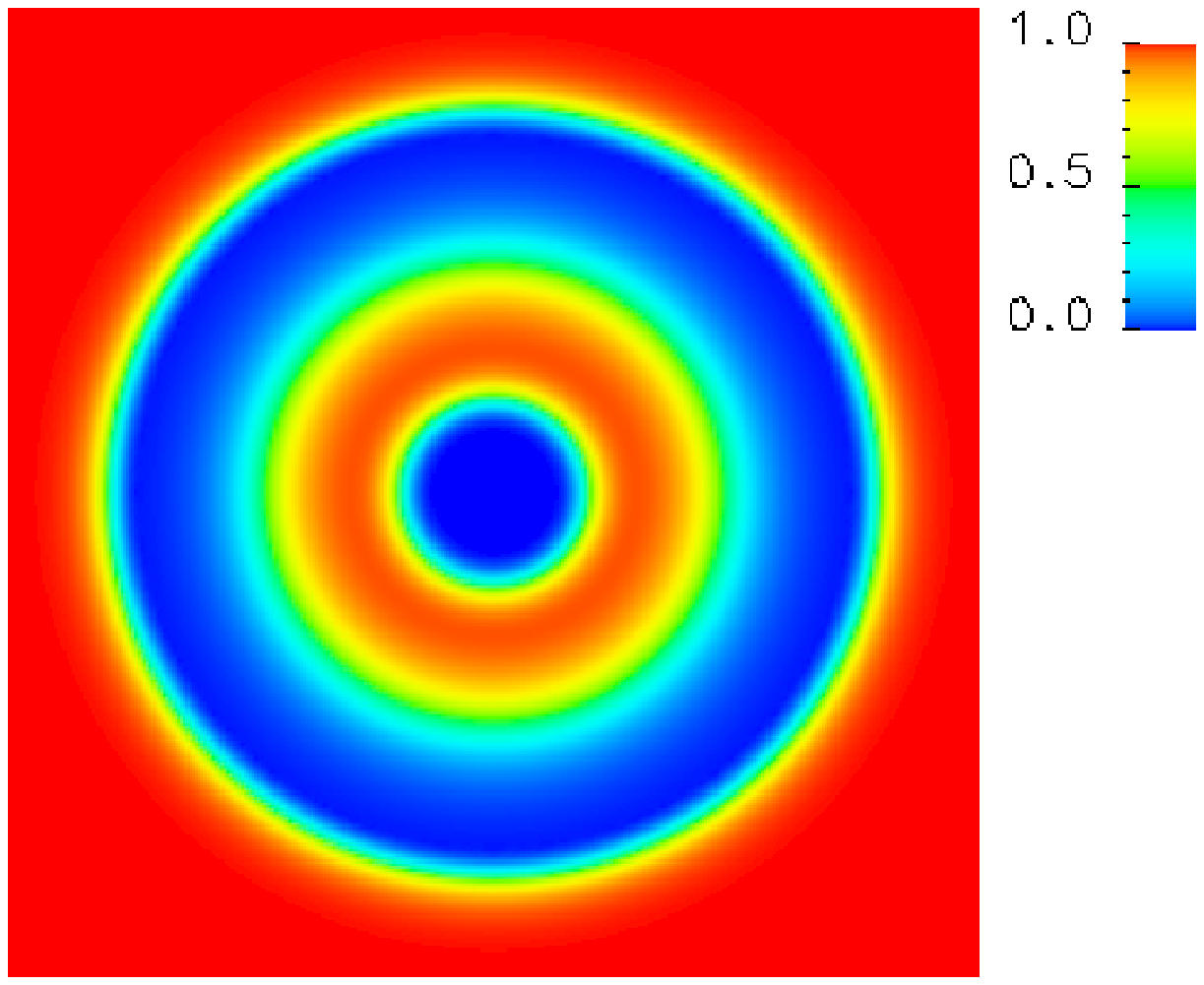}
\includegraphics[width=8cm]{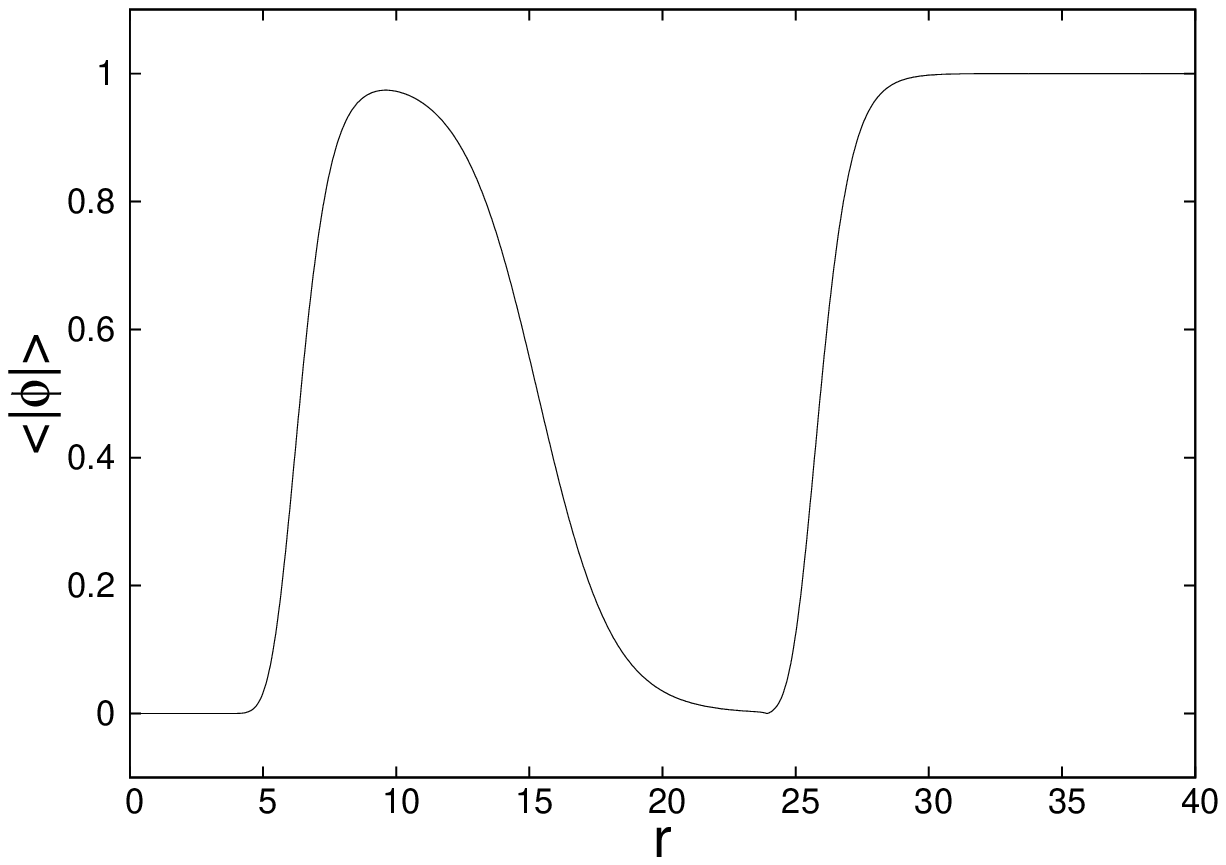}
\caption{
A two-layer solution in which the inner layer has vortex number 
$N_1=10^2$ with bag parameter $p_1=0$ and the outer layer has vortex number 
$N_2=10^8$ with bag parameter $p_2=0.3.$
The left-hand image displays the modulus of the Higgs field $|\phi|.$ 
The right-hand plot is the angular average $\langle|\phi|\rangle$
 as a function of the radius $r.$  
}
\label{fig-ml2}\end{center}\end{figure}

\section{Multi-layer magnetic bags}\news
Manton \cite{Ma} has applied a generalization of the magnetic bag 
approximation to study monopoles with a multi-layer structure,
consisting of a sequence of magnetic bags with increasing magnetic 
charges. Multi-layer magnetic bags for hyperbolic vortices
can be obtained as exact BPS solutions by taking $f(z)$ to have the
obvious product form
\be
f(z)=z\prod_{j=1}^{M}\frac{z^{N_j}-1+N_j^{-p_j}}{1-(1-N_j^{-p_j})z^{N_j}}.
\label{ml}
\ee
For appropriate values of $N_j$ and $p_j,$ with $j=1,\ldots,M,$
this solution describes a multi-layer configuration with $M$ layers.
Only the innermost layer can be a core magnetic bag, because
the zeros of the Higgs field must be at the origin in this case.
If two layers are core magnetic bags, say $p_k=p_{k+1}=0,$ then 
clearly the two layers degenerate to a single core magnetic bag 
with vortex number $N_k+N_{k+1}.$

As illustration, $|\phi|$ is displayed in Figure~\ref{fig-ml2} for
a two-layer solution with $N_1=10^2, \ p_1=0$ and $N_2=10^8, \  p_2=0.3.$ 
The two-layer structure is clearly visible in this plot, as is the behaviour
of the Higgs field, which is close to zero in each bag but returns to
approximately its vacuum expectation value between the layers.  
Also plotted in Figure~\ref{fig-ml2} is the angular average 
$\langle|\phi|\rangle$ as a function of the radius $r,$ displaying 
transitions between regions where the Higgs field is close to zero and
regions where it is close to its vacuum expectation value.  

The angular average confirms that this type of solution has approximate 
circular symmetry and can therefore be well-approximated by a 
multi-layer magnetic bag description that assumes this symmetry.
The appropriate form of this type of multi-layer magnetic bag description is
clearly a set of concentric non-overlapping annuli, inside which the
Higgs field is assumed to vanish and outside which it is taken to
have its vacuum expectation value. Within this approximation, the 
vortex number in each layer is equal to the area of the annulus 
divided by $4\pi.$ Obviously, the vortex number is higher in layers that
are further out and, in particular, to have equally spaced layers requires
an exponential growth of the vortex number. The innermost annulus can
degenerate to a disc, to describe a core magnetic bag at the centre, but
this is the only annulus that is allowed to degenerate because of the
non-overlapping requirement.  

\begin{figure}[ht]\begin{center}
\includegraphics[width=8cm]{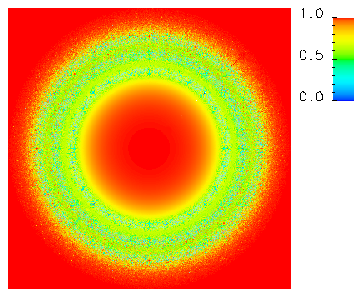}
\includegraphics[width=8cm]{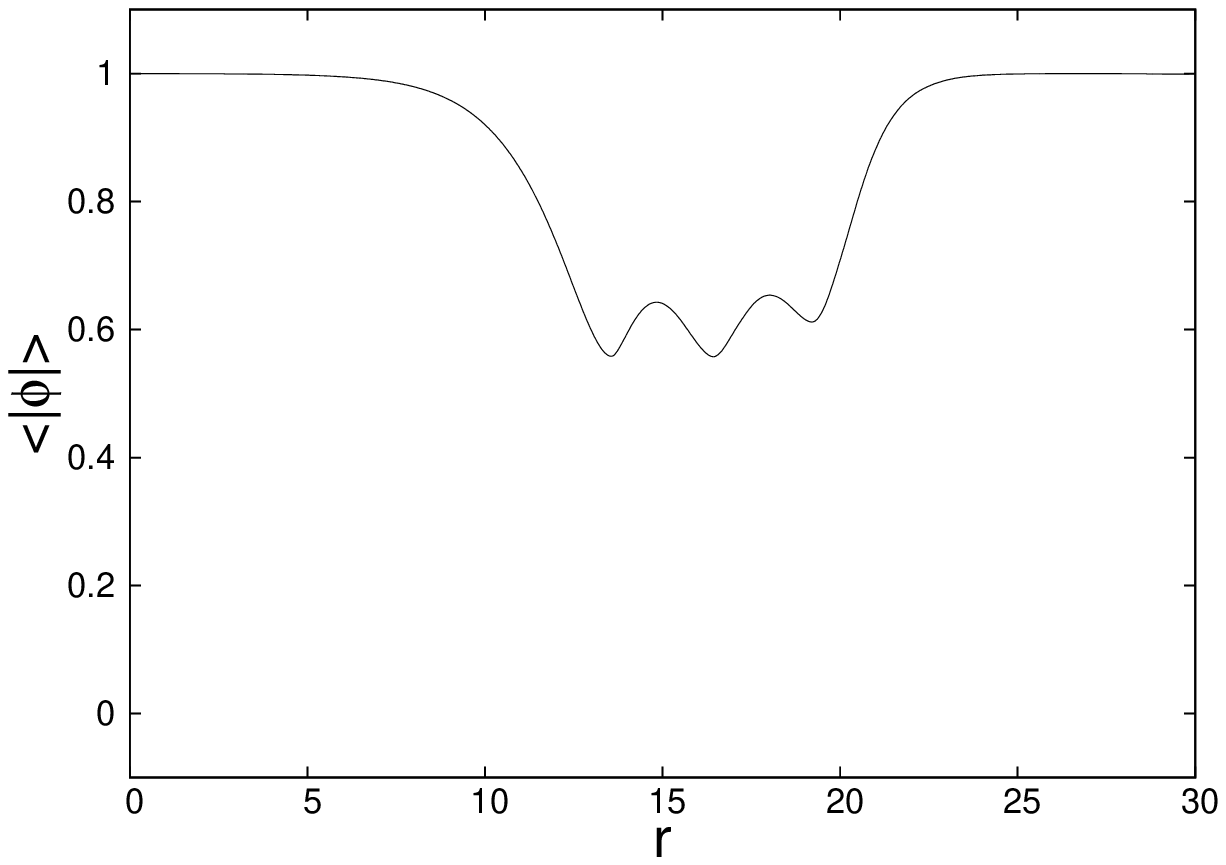}
\caption{
A three-layer solution in which the inner layer has vortex number 
$N_1=10^4$ with bag parameter $p_1=1.0,$ 
the middle layer has 
$N_2=10^5$ with $p_2=0.9$ and the 
outer layer has $N_3=10^6$ with $p_3=0.8.$
The left-hand image displays the modulus of the Higgs field $|\phi|.$ 
The right-hand plot is the 
angular average $\langle|\phi|\rangle$ as a function of the radius $r.$  
}
\label{fig-ml3}\end{center}\end{figure}

The description presented above for hyperbolic vortices contrasts 
with the behaviour of the Higgs field predicted in \cite{Ma} 
for multi-layer monopoles using the magnetic bag approximation.
In the monopole case, the modulus of the assumed abelianized
Higgs field is not close to zero in each layer but rather approaches an 
effective vacuum expectation value determined by the magnetic charges
and radii of the multi-layer system. Given the above discussion, it is
clear that this description can only apply to the angular average of
the Higgs field in a dilute regime. Indeed the description in \cite{Ma}
is of multi-layers consisting of dilute spherical clusters of large 
numbers of monopoles. This is consistent with the analysis in \cite{LW},
where it is observed that $N$ monopoles located at the bag surface 
(which has radius $O(N)$) have an inter-monopole separation that is
$O(\sqrt{N}).$
 
To produce a hyperbolic vortex analogue of 
the kind of multi-layer structure envisaged for monopoles, requires 
moving to the dilute regime, with bag deformation parameters closer
to unity than zero. An example is presented in Figure~\ref{fig-ml3},
for a three-layer solution where the vortex numbers
are $N_1=10^4,\ N_2=10^5,\ N_3=10^6$ and the bag parameters are
given by $p_1=1.0,\  p_2=0.9,\  p_3=0.8.$ The dilute nature of the layers
is clearly visible in this plot and contrasts with the dense regime
displayed previously in Figure~\ref{fig-ml2}. The vortex numbers and
bag parameters have been chosen so that the layers are close together
and produce a large region where the angular average of the modulus 
of Higgs field is reasonably constant but at a value that is only 
about half of the Higgs expectation value. 
A plot of the angular average $\langle |\phi|\rangle$ as a function of
the radius is included in Figure~\ref{fig-ml3} to demonstrate this
feature.

To provide a magnetic bag description of a dilute solution
requires working with average values, as follows. Consider a region with
area ${\cal A}$ and let $\varphi^2$ be the average value of $|\phi|^2$
throughout this region. The appropriate magnetic bag approximation is to
assume that $|\phi|$ takes its vacuum expectation value outside this 
region and to average the second Bogomolny equation (\ref{bog}) inside 
the region to provide the approximate vortex number as 
$N={\cal A}(1-\varphi^2)/(4\pi).$ This formula shows that
the previous dense magnetic bag approximation, which corresponds to the 
choice $\varphi^2=0,$ yields minimal area for fixed vortex number.
This mimics the interpretation of the non-abelian monopole magnetic
bag as the most compact arrangement of $N$ monopoles.

The two examples displayed in  
Figure~\ref{fig-ml2} and Figure~\ref{fig-ml3}
are simply representative solutions that highlight the type of behaviour
that arises.
The solution presented in Figure~\ref{fig-ml3} does not contain a 
core magnetic bag at its interior, but this could be added by including
an additional layer with a vanishing bag parameter. Indeed it should
be clear that there are many possible variations on multi-layer magnetic
bags that may or may not have core magnetic bags at their centre and can
include combinations of both dense and dilute layers. 
In the continuum limit this allows the construction of hyperbolic
 vortex analogues of Manton's monopole planets and galaxies \cite{Ma}.
 
\section{Conclusion}\news
Exact BPS vortex solutions of the abelian Higgs model in the hyperbolic plane
have been used to provide an explicit realization of magnetic bags.
This has allowed an investigation of some of their properties and 
associated phenomena predicted in the context of BPS monopoles,
including different extreme types of bags and multi-layer bags.

Witten's ansatz \cite{Wi2} provides a direct mapping between the solutions
considered in this paper and $SU(2)$ Yang-Mills instantons in 
four-dimensional Euclidean space with an $SO(3)$ rotational symmetry.
The instanton number is identified with the vortex number $N,$ so 
magnetic bags for vortices map to instantons with a large instanton
number $N\gg 1$. It might be interesting to follow through the details
of this correspondence and in particular to see how these instantons
are described within a large $N$ limit of the ADHM construction \cite{ADHM},
in a manner similar to the recent study \cite{Ha} of the large $N$ limit of the 
Nahm transform and its relation to monopole magnetic bags.  

\section*{Acknowledgements}
Many thanks to Nick Manton for useful comments.
I acknowledge EPSRC and STFC for grant support.

\end{document}